# Can social media provide early warning of retraction? Evidence from critical tweets identified by human annotation and large language models

This is the preprint version of a paper that has been published in the *Journal of the Association for Information Science and Technology*.

***Cite as***: Zheng, E.-T., Fu, H.-Z., Thelwall, M., & Fang, Z. (2025). Can social media provide early warning of retraction? Evidence from critical tweets identified by human annotation and large language models. *Journal of the Association for Information Science and Technology*. https://doi.org/10.1002/asi.70028

Er-Te Zheng[1], Hui-Zhen Fu[2], Mike Thelwall[1], Zhichao Fang[3,4*]

*Corresponding author: z.fang@cwts.leidenuniv.nl

1. School of Information, Journalism and Communication, The University of Sheffield, Sheffield, The UK.

2. Department of Information Resources Management, Zhejiang University, Hangzhou, China.

3. School of Information Resource Management, Renmin University of China, Beijing, China.

4. Centre for Science and Technology Studies (CWTS), Leiden University, Leiden, The Netherlands.

### Abstract

Timely detection of problematic research is essential for safeguarding scientific integrity. To explore whether social media commentary can serve as an early indicator of potentially problematic articles, this study analysed 3,815 tweets referencing 604 retracted articles and 3,373 tweets referencing 668 comparable non-retracted articles. Tweets critical of the articles were identified through both human annotation and large language models (LLMs). Human annotation revealed that 8.3% of retracted articles were associated with at least one critical tweet prior to retraction, compared to only 1.5% of non-retracted articles, highlighting the potential of tweets as early warning signals of retraction. However, critical tweets identified by LLMs (GPT-4o mini, Gemini 2.0 Flash-Lite, and Claude 3.5 Haiku) only partially aligned with human annotation, suggesting that fully automated monitoring of post-publication discourse should be applied with caution. A human-AI collaborative approach may offer a more reliable and scalable alternative, with human expertise helping to filter out tweets critical of issues unrelated to the research integrity of the articles. Overall, this study provides insights into how social media signals, combined with generative AI technologies, may support efforts to strengthen research integrity.

## 1. Introduction

*Problematic articles* are scholarly publications containing methodological flaws, data irregularities, or ethical violations, regardless of whether these issues have been formally identified by the scientific community. The presence of such articles can mislead researchers, policymakers, and the public (Candal-Pedreira et al., 2022; Larsson, 1995). When concerns identified after publication are deemed sufficiently serious, problematic articles may ultimately be retracted. Retraction serves as the primary corrective mechanism, defined as “a mechanism for correcting the literature and alerting readers to articles that contain such seriously flawed or erroneous content or data that their findings and conclusions cannot be relied upon” (COPE Council, 2019). *Retracted articles* are thus outputs that have been officially withdrawn by journals or publishers due to their unreliability or other concerns.

Despite a notable increase in retractions in recent years (Van Noorden, 2011, 2023; Vuong et al., 2020; Zheng et al., 2025), the retraction process often remains slow and opaque. Problematic articles can remain accessible and continue exerting influence – sometimes accompanied only by a statement of concern – long before decisive editorial action is taken. This delay underscores the need for earlier detection of problematic content and mitigation of its potential harms (Bar-Ilan & Halevi, 2017).

Beyond pre-publication peer review, traditional methods for identifying problematic articles have largely focused on detecting plagiarism (Eysenbach, 2000; Wager, 2011) and image manipulation (Bik et al., 2016; Koppers et al., 2017). However, these approaches have limited scope and less effective in detecting other forms of misconduct, such as data fabrication, falsification, or authorship disputes. In this context, social media is increasingly recognised as a component of post-publication peer review (Irawan et al., 2024).

Although criticism on platforms such as Twitter[1], Facebook, or academic blogs does not inherently indicate that an article is problematic or destined for retraction, heightened public scrutiny may prompt further investigation by journals, institutions, or the broader scientific community. In some cases, such scrutiny has accelerated editorial actions, including retractions. Notable examples include exposure of papers containing AI-generated text (Tarla et al., 2023; M. Zhang et al., 2024), manipulated images (X. Guo et al., 2024; Wu et al., 2024), or high-profile studies – such as those related to COVID-19 – that were criticised for lacking robust empirical support (Mehra et al., 2020; Walach et al., 2021).

[1] Twitter was the official name of X at the time of data collection, so we use the terms “Twitter” and “tweets” in this study.

These observations suggest that critical commentary on social media may serve as an early warning signal for identifying problematic articles before formal retraction. While social media discourse has been widely studied within the framework of *altmetrics* as a measure of attention surrounding scholarly publications, it has not yet been systematically explored as a tool for surfacing concerns about scientific integrity.

### *1.1 Altmetric research on retracted articles*

Existing altmetric research has shown that retracted articles often attract considerable attention on social media, especially on platforms such as Twitter and among the general public (Bornmann & Haunschild, 2018; Khademizadeh et al., 2024; Khan et al., 2022). A positive correlation has been observed between high Altmetric Attention Scores (AAS) and the likelihood of retraction due to misconduct (Shema et al., 2019), suggesting that social media visibility may signal underlying issues.

Moreover, retracted articles typically receive more social media engagement than comparable non-retracted articles (Peng et al., 2022; Serghiou et al., 2021; Sotudeh et al., 2022). Metrics such as the number of tweets, retweets, likes, and replies are often higher for retracted articles, indicating greater public scrutiny or interest (Dambanemuya et al., 2024). Most of this engagement occurs before a formal retraction notice is issued (Dambanemuya et al., 2024; Serghiou et al., 2021). This suggests that monitoring social media activity may help identify problematic content at an earlier stage.

Beyond volume, several studies have examined the content of tweets referencing retracted articles. Dambanemuya et al. (2024) found that retracted articles attracted significantly more tweets containing retraction-related keywords than non-retracted ones, both before and after retraction. Similarly, Peng et al. (2022) reported that tweets about retracted articles were more likely to include critical commentary. Sentiment analysis further supports this: approximately one-third of retracted articles are associated with critical sentiment in tweets both pre- and post-retraction, with critical tweets retweeted more frequently than positive or neutral ones (Amiri et al., 2024).

Case-based evidence reinforces this trend. For example, Haunschild and Bornmann (2021) found that critical tweets questioning the validity of two out of three COVID-19 articles began circulating shortly after their publication. Despite these insights, no prior research appears to have systematically leveraged social media content to assess whether an article is likely to be problematic or at risk of retraction.

### *1.2 Patterns of Twitter engagement with scholarly articles*

Mentions of scholarly articles on Twitter constitute one of the most prevalent forms of altmetric data (Costas et al., 2015; Ortega, 2018; Sugimoto et al., 2017; Thelwall et al., 2013). While both scientists and the general public contribute to the dissemination of research on Twitter (Yu et al., 2019; L. Zhang et al., 2023), scientists remain the primary participants in science-related discussions (Carlson & Harris, 2020). They are more likely to use the platform to communicate with colleagues, establish collaborations, engage in advocacy or political debates, share findings with broader audiences, and stay informed about developments in their fields (Bowman, 2015). The public, in turn, often follows scientists and generally responds positively to scientific engagement on social media (Côté & Darling, 2018). Notably, the topics attracting the attention of scientists and the public tend to be closely aligned (Karmakar et al., 2023).

Twitter engagement typically peaks shortly after an article's publication, making tweet counts one of the fastest-emerging forms of altmetric data (Fang & Costas, 2020; Yu et al., 2017). Around 40% of article-related tweets appear within the first week (Priem & Costello, 2010). However, most interactions consist of low-effort engagement, such as likes and retweets. Higher-level engagement – defined as actions involving discussion, interpretation, critique, or questioning of content – accounts for only a small portion of interactions (Z. Fang et al., 2022). In certain cases, particularly those involving controversial or socially salient topics such as climate change or COVID-19, scholarly tweets have sparked broader public debates (Alperin et al., 2024; Toupin et al., 2022).

Users' motivations for tweeting about scholarly articles are diverse, encompassing scientific information seeking and sharing, professional goals such as network-building or institutional promotion, and communicative purposes like explaining findings to peers or lay audiences (Mohammadi et al., 2018). As such, Twitter plays a multifaceted role in scientific dissemination (Chatterjee et al., 2020), including informal post-publication peer review. Similarly to negative citations in the scholarly literature, tweets can take a critical tone – expressing scepticism, methodological concerns, or even accusations of misconduct (Teixeira da Silva & Dobránszki, 2019). In some of the aforementioned cases, such discussions have brought serious errors or ethical breaches to light.

### *1.3 Large language models for classification tasks*

Large language models (LLMs), such as ChatGPT, Claude, and Gemini, are transformer-based architectures trained on vast text corpora, enabling them to perform a wide range of natural language processing (NLP) tasks (Zubiaga, 2024). One such task is text classification, defined as the assignment of predefined labels to text, such as sentiment or topic. The ability of LLMs to generalise across domains through in-context learning and prompt engineering makes them particularly promising for classification tasks (Vajjala & Shimangaud, 2025; Wang et al., 2024).

Empirical studies have shown that LLMs often outperform traditional machine learning methods in both multiclass and binary classification tasks. For instance, LLMs have achieved higher accuracy in predicting employees' work locations based on job reviews and in classifying news articles as real or fake (Kostina et al., 2025). In zero-shot settings, where no task-specific training data or examples are provided, LLMs have surpassed support vector machines and even pretrained transformer-based models such as RoBERTa in recall, thereby reducing false negatives during preprocessing (Y. Guo et al., 2024). ChatGPT has also outperformed crowd workers in various annotation tasks, including relevance, stance, topic, and frame detection for tweets and news articles, with average zero-shot accuracy gains of around 25 percentage points (Gilardi et al., 2023). In addition, LLMs have demonstrated substantial agreement with human annotators in taxonomic labelling across diverse social science corpora (Ziems et al., 2024).

Despite these advances, LLMs have important limitations. They are prone to hallucinations, generating inaccurate or fabricated content that deviates from factual knowledge or the training data (Perković et al., 2024). They may also replicate demographic and historical biases embedded in their training corpora, potentially amplifying stereotypes and producing unfair outputs (Ranjan et al., 2024). Another constraint is the limited context window of most models, which restricts their ability to process long documents. Although recent improvements have extended input capacity, it remains a performance bottleneck for tasks requiring comprehensive document comprehension (Hosseini et al., 2025).

In industry, LLMs have recently been adopted to evaluate social media engagement with scholarly content. For example, Altmetric has introduced a beta feature that applies LLM-powered sentiment analysis to posts on X (formerly Twitter) and Bluesky.[2] This tool classifies posts into seven categories – ranging from strong negative to strong positive – in an effort to capture more nuanced public discourse about research. While this represents a practical step forward, it remains unclear how effectively LLMs can detect *critical content* specifically directed at scholarly articles. If LLMs can reliably identify such criticism, they may offer a scalable method for early detection of potentially problematic research, thereby contributing to efforts to uphold research integrity.

### *1.4 Objectives of this study*

The widespread engagement with retracted articles on Twitter, along with the growing application of LLMs for sentiment analysis by platforms such as Altmetric, highlights the need to assess whether social media commentary can serve as an early warning signal for problematic research, and whether LLMs can support this effort.

This study addresses these questions in two parts. First, we examine whether retracted articles attract more critical tweets *prior to* retraction than non-retracted articles, based on human annotation. *Critical tweets* are operationally defined as posts that express criticism, accusations, doubts, sarcasm, irony, or mockery directed at a scholarly article – a definition partially overlapping with Altmetric's negative sentiment categories.

Second, we evaluate the accuracy of LLMs in identifying critical content in tweets, using human annotation as the benchmark. This evaluation not only provides insights into the reliability of

[2] For more information on the sentiment analysis introduced by Altmetric, see: https://help.altmetric.com/support/solutions/articles/6000279392-sentiment-analysis-in-altmetric (accessed on 19 June 2025).

LLM-powered sentiment systems for detecting potentially problematic research, but also contributes empirical evidence to support or challenge the validity of using LLMs in the analysis of science communication.

Specifically, we address the following research questions (RQs):

- **RQ1**. To what extent do critical tweets appear in Twitter discussions of retracted versus non-retracted articles?
- **RQ2**. How accurately can LLMs detect whether a tweet about a scholarly article is critical of it, using human annotation as the benchmark?

## 2. Conceptual framework

Scholarly peer review is widely recognised as an imperfect process (Heesen & Bright, 2021). Many articles are retracted post-publication despite having passed peer review, leading to wasted research resources – especially for those building upon or attempting to replicate flawed findings – and ultimately undermining trust in the scientific enterprise (Peterson, 2018). In response to concerns about articles that have not been retracted, a complementary evaluation mechanism has emerged: post-publication peer review (PPPR).

Platforms such as PubPeer offer dedicated spaces for ongoing critical discussion of published academic work. These forums often help to publicise methodological or factual errors and facilitate constructive critique, which can contribute to article retractions (Bordignon, 2020). However, relatively few researchers actively participate on formal PPPR platforms. Writing thoughtful and methodologically sound critiques is time-consuming, and early-career researchers may hesitate to publicly criticise senior academics due to fear of reputational or professional repercussions (Bastian, 2014).

Informal PPPR has long existed within academia, for example through discussions at conferences or within research groups (Peterson, 2018). With the rise of social media as a tool for informal scholarly communication, these discussions also occur online and can be conceptualised as a form of PPPR (Irawan et al., 2024). Compared to formal platforms, social media offers several unique advantages: its brevity lowers participation barriers, and the possibility of anonymity may encourage researchers – especially junior ones – to voice criticism without fear of professional consequences. Twitter, in particular, plays a central role, accounting for perhaps 80% of social media discussions about scholarly publications (Peng et al., 2022).

Despite its accessibility, the reliability of social media commentary is variable. Users without domain expertise may post unsubstantiated, irrelevant, or misleading content (Ali & Watson, 2016). Unlike specialised PPPR platforms, which often feature detailed critiques that can directly lead to retractions, the role of social media in influencing retraction decisions has yet to be systematically studied at scale. Given Twitter's prominence in science communication, this study focuses on critical tweets as a lens for examining how social media may contribute to the retraction process.

We conceptualise tweets as playing a dual role in the post-publication lifecycle of scholarly articles (see Figure 1), corresponding to two distinct yet interconnected influence pathways through which critical Twitter discussions may affect retraction outcomes:

- *Instrumental pathway*: Tweets that explicitly identify methodological flaws, fabricated data, or ethical violations may function as whistle-blowing signals. Once detected – whether by experts, editorial monitoring tools, or through direct tagging of editors or journals – such tweets may trigger formal investigations by journals or institutions.
- *Social amplification pathway*: Drawing on the *social amplification of risk framework* (Kasperson et al., 1988), initial critical tweets may be retweeted and discussed by others, extending their reach beyond the original audience. Over time, scholars, influencers, or topic-focused accounts may reframe and reinforce the concerns. This cascade of attention can eventually involve mainstream media or influential blogs, amplifying perceived risks and potentially exerting indirect pressure on journals or funders to take formal action, sometimes culminating in article retraction.

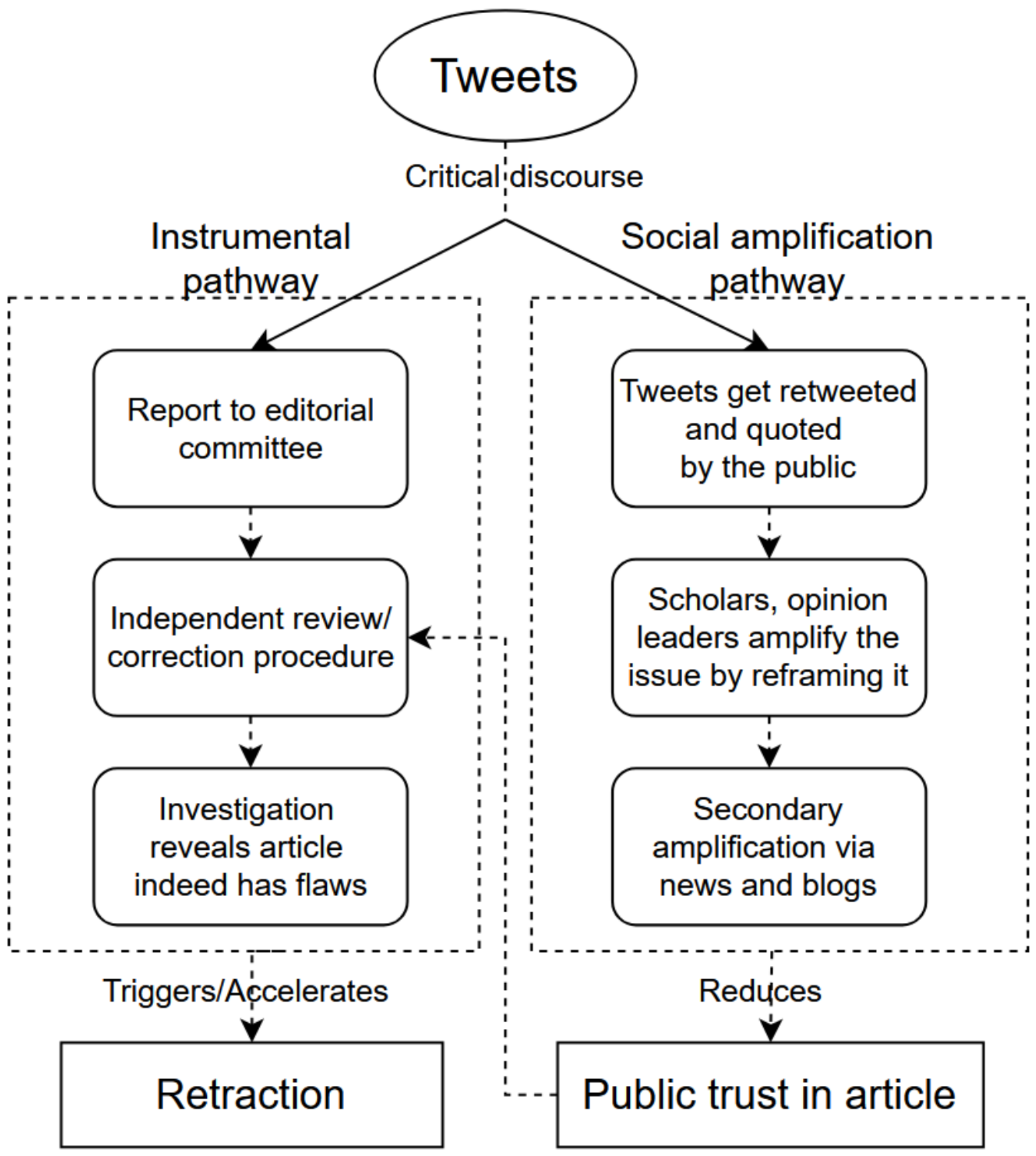

**Figure 1**. Conceptual model of Twitter's influence pathways in the post-publication peer review (PPPR) process of problematic articles.

## 3. Data and methods

*3.1 Dataset of retracted articles*

To identify retracted articles, we used a snapshot of the Web of Science (WoS) database (version dated March 2025), maintained by the Centre for Science and Technology Studies (CWTS) at Leiden University. From the WoS *Core Collection*, we selected articles published in 2019 that were classified as "Retracted Publication" under the document type field. The publication year 2019 was chosen to avoid anomalies in retraction patterns associated with the surge of COVID-19-related publishing in subsequent years. This process yielded a dataset of 2,387 distinct retracted articles, for which bibliographic metadata were also extracted from the WoS database. To determine the exact retraction dates, we consulted *Retraction Watch* (accessed in June 2025), which serves as the most comprehensive repository of retracted articles (Brainard, 2018).

Using DOI-based searches in the Altmetric database (snapshot dated November 2022), also maintained by CWTS, we retrieved the tweet IDs of original tweets referencing these retracted articles, as recorded by Altmetric up to that date. The corresponding tweet content was then retrieved via the Twitter API in March 2023,[3] based on the retrieved tweet IDs. The resulting dataset included each tweet's full text and timestamp (i.e., date and time of posting).

As this study focuses specifically on *pre-retraction* discourse, we excluded any retracted articles that lacked original tweets posted before their official retraction date. This filtering step resulted in a dataset of 712 retracted articles that had at least one pre-retraction tweet, representing 29.8% of the initial sample.

*3.2 Methods*

Figure 2 illustrates the workflow of data sampling, cleaning, and analysis. Each methodological step is described in detail below.

[3] Following the acquisition of Twitter by Elon Musk and its subsequent rebranding as X, the platform discontinued free access to its API in March 2023. Consequently, the tweet data collected in March 2023 represents the most recent large-scale dataset obtainable under the previous access conditions. The lack of up-to-date tweet data constitutes a key limitation of this study. However, since the majority of tweets referencing scholarly articles typically accumulate shortly after publication (Z. Fang & Costas, 2020; Priem & Costello, 2010; Yu et al., 2017), the four-year window (2019-2022) is likely sufficient for capturing most Twitter engagement with the articles published in 2019 that are included in our datasets. Additionally, given that Twitter (now X) remains the most prominent source of altmetric data, the methodology developed in this study remains applicable to scholarly discourse on the platform and is potentially transferable to other similar social media environments.

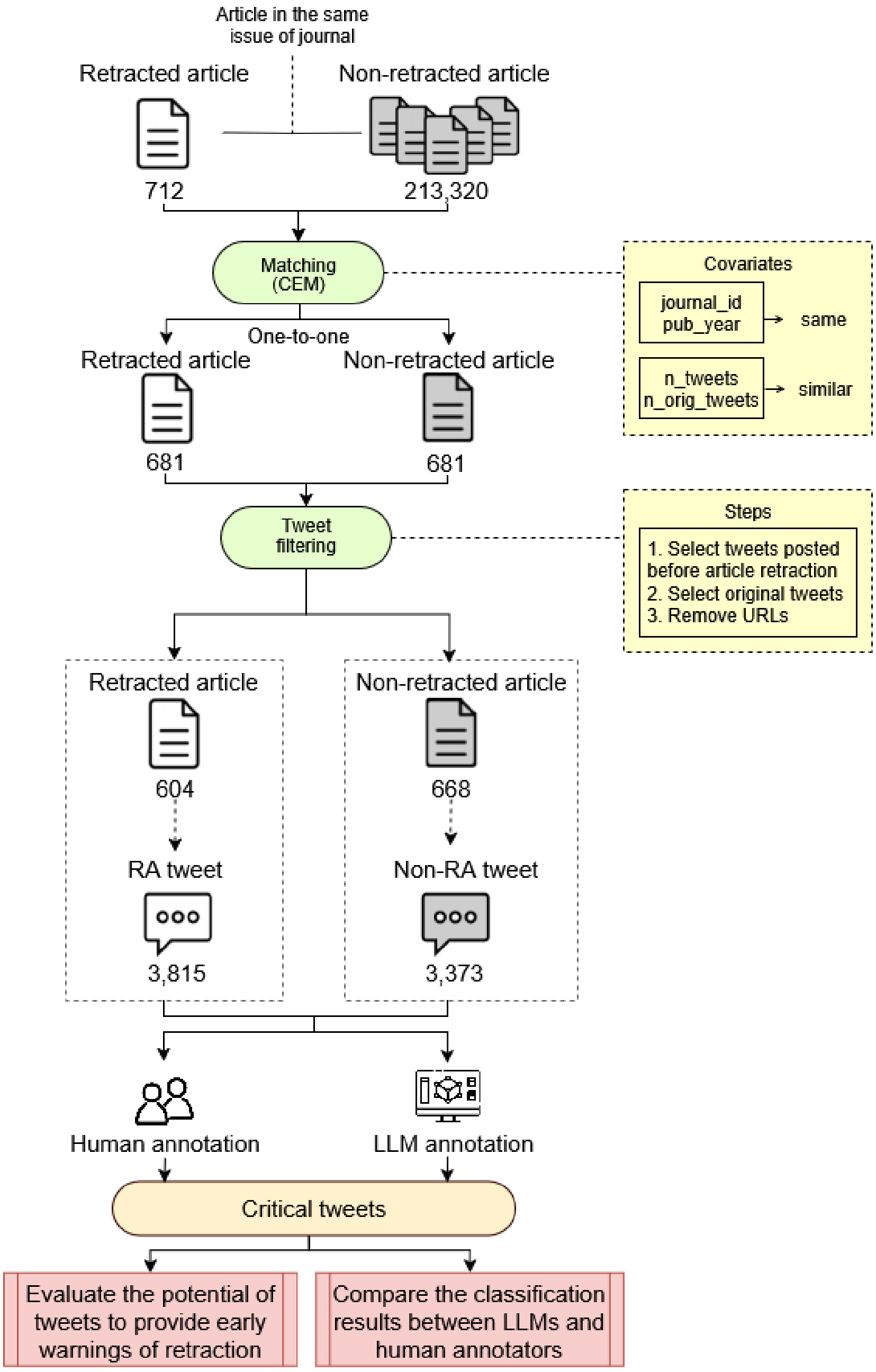

**Figure 2**. Research workflow of the study. *Note*: "RA" refers to "retracted articles" and "non-RA" to "non-retracted articles", both here and throughout the text.

(1) Matching method: Coarsened Exact Matching (CEM)

To assess whether Twitter discussions can serve as early warning signals for articles that are subsequently retracted, we analysed tweets referencing either retracted or non-retracted articles (i.e., those not retracted at the time of data collection). Coarsened Exact Matching (CEM) was applied to construct a comparison group of non-retracted articles matched to those in the retracted dataset. CEM is a matching technique that involves coarsening the values of covariates into groups and then performing exact matching within each group (Blackwell et al., 2009;

Iacus et al., 2012). This method reduces the imbalance of covariates between treatment and control groups by excluding observations that lack suitable matches.

Non-retracted articles were drawn from the WoS database and matched to retracted articles based on four covariates: (1) publication venue (i.e., journal), (2) publication year, (3) total number of tweets received, and (4) number of original tweets (excluding retweets). This ensured that the matched retracted and non-retracted articles were published in the same journal and year (2019) and had comparable levels of Twitter attention. Of the 591,527 non-retracted articles published in the same journal issues in 2019 as the 712 retracted articles in our dataset, 213,320 had at least one tweet recorded by Altmetric.

After applying CEM, a total of 1,362 articles were successfully matched, comprising 681 retracted and 681 non-retracted articles. Bibliographic and Twitter data for the non-retracted articles were collected using the same procedures as for the retracted articles.

(2) Tweet filtering

We refined the dataset through three filtering steps: First, for retracted articles, only tweets posted before the retraction date were retained. Second, retweets were excluded to ensure that all analysed tweets contained original textual content. Third, URLs were removed from tweet texts to focus exclusively on semantic content. The final dataset comprised 3,815 tweets associated with 604 retracted articles and 3,373 tweets associated with 668 comparable non-retracted articles.

(3) Baseline: human annotation

Two authors (Zheng and Fang) independently annotated all 7,188 tweets to determine whether they expressed criticism of the referenced article (see Table 1). Based solely on tweet text, tweets were annotated as *critical* if they conveyed criticism, accusations, doubts, sarcasm, irony, or mockery directed at the article – features considered indicative of potential issues with the article. Tweets meeting this criterion were assigned a label of 1 (*critical*); otherwise, they were labelled 0 (*non-critical*). To ensure consistency, the annotation process followed predefined criteria, and any uncertainties were flagged for subsequent discussion and resolution.

**Table 1.** Confusion matrix of human annotation by the two annotators.

| | Fang: Non-critical | Fang: Critical |
|---|---|---|
| Zheng: Non-critical | 6,785 | 71 |
| Zheng: Critical | 154 | 178 |

To evaluate inter-coder reliability, Gwet's AC1 coefficient was calculated. This chance-adjusted metric offers greater stability than Cohen's Kappa (Wongpakaran et al., 2013). The resulting Gwet's AC1 value of 0.969 indicates a high level of agreement. Remaining discrepancies were resolved through discussion to finalise the annotation. Table 2 reports examples of tweets annotated as critical, illustrating the range of rhetorical styles included in the classification.

**Table 2.** Examples of tweets annotated as critical by the two human annotators.

| Category | Example tweet |
| --- | --- |
| Criticism, accusations | *"This paper's methodology is seriously flawed, they cherry-picked data to fit their narrative."* |
| Doubts | *"I bet the authors fudged the results to get published in that journal, smells fishy."* |
| Sarcasm, irony, or mockery | *"Impressive! Anyone can get into PubMed these days by tossing a draft to some shady open-access journal."* |

*Note:* Tweet content has been paraphrased to protect user privacy.

(4) Large language models (LLMs)

Three LLMs – GPT-4o mini, Gemini 2.0 Flash-Lite, and Claude 3.5 Haiku – were employed to classify tweets as either critical or non-critical. These transformer-based models represent leading LLMs for natural language understanding tasks (Akpan, 2025). The selected model versions were chosen for their high performance-to-cost ratio, making them well-suited for tasks that require nuanced language comprehension under moderate computational constraints.

For each LLM, we conducted experiments using both zero-shot and few-shot prompting strategies (Figure 3). In zero-shot prompting, the model performs the task without being given any exemplars. In contrast, few-shot prompting provides the model with a small set of illustrative examples embedded in the prompt to guide its output (Brown et al., 2020). Employing both strategies allowed for a systematic evaluation of model performance with and without exemplar-based guidance.

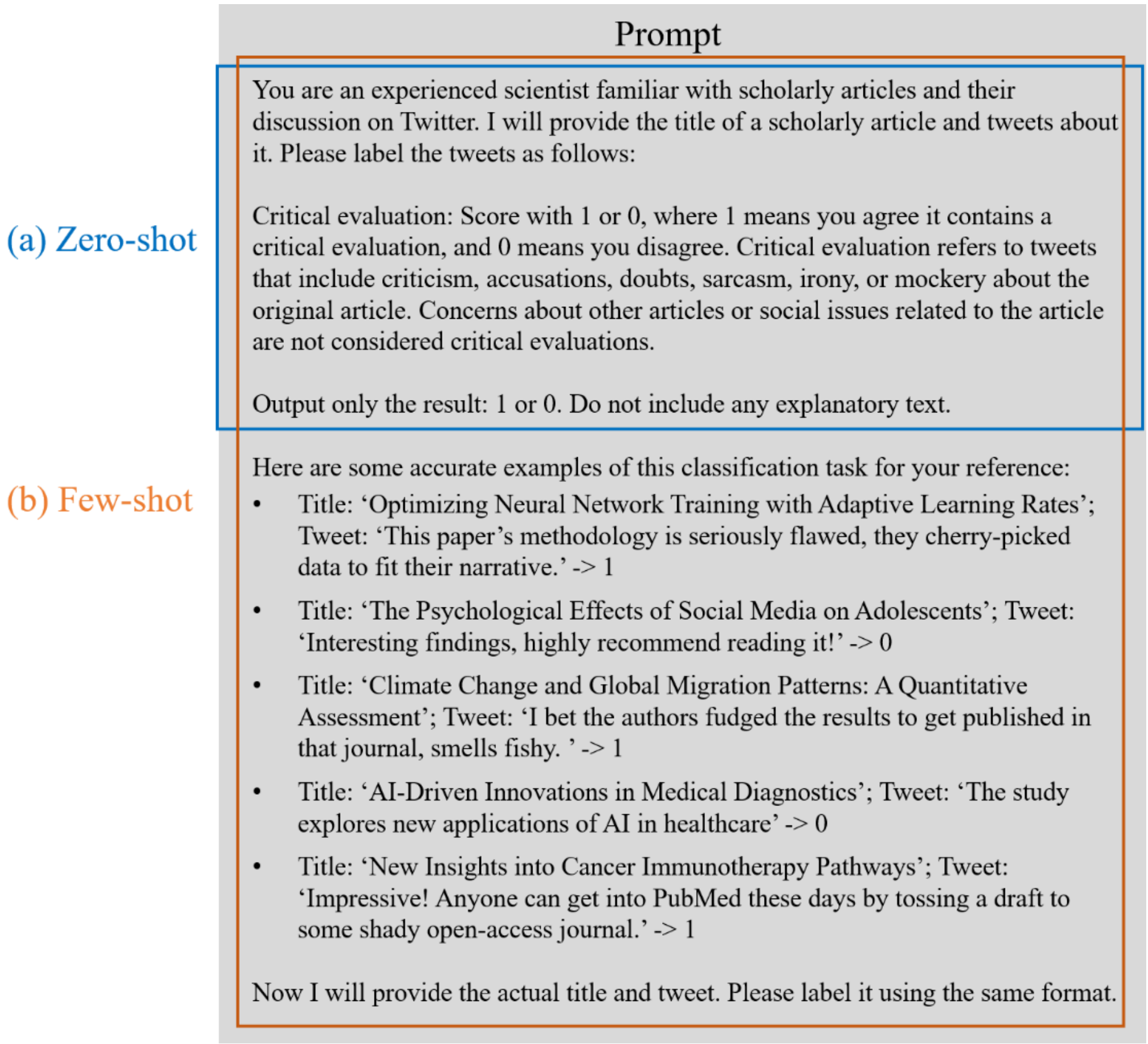

**Figure 3.** Prompt settings used for the LLMs: (a) zero-shot and (b) few-shot.

All LLM runs were conducted via their respective APIs[4], with the temperature parameter set to 0 to maximise output consistency and reproducibility. Experiments were performed on 10 June 2025. For each tweet, three independent classification runs were executed per model under both zero-shot and few-shot prompt settings. The majority vote across the three runs was recorded as the model's consensus output for each prompt setting, which was then compared against the human-annotated ground truth. In addition to individual model evaluations, we generated a cross-model consensus output – defined as the majority vote across the consensus outputs of all three LLMs – to assess whether ensemble agreement yielded better alignment with human annotation than any single model.

The consistency rate – defined as the proportion of identical classification outputs across repeated runs of the same model – is reported in Table 3. The results show high internal consistency across all models and prompt settings, suggesting that the models produced highly reproducible outputs under fixed conditions.

[4] ChatGPT: https://platform.openai.com/docs/api-reference/assistants/createAssistant (accessed on 31 March 2025); Gemini: https://ai.google.dev/gemini-api/docs/models/generative-models?hl=en (accessed on 31 March 2025); Claude: https://docs.anthropic.com/en/api/complete (accessed on 31 March 2025).

**Table 3.** Consistency rates across repeated runs for each model.

| Prompt setting | GPT-4o mini | Gemini 2.0 Flash-Lite | Claude 3.5 Haiku |
|---|---|---|---|
| Zero-shot | 98.8% | 99.9% | 99.6% |
| Few-shot | 99.4% | 99.8% | 99.5% |

*3.3 Indicators*

To evaluate the performance of the LLMs in identifying tweets containing critical commentary, we employed three widely used evaluation metrics: precision, recall, and F1-Score. These indicators offer complementary perspectives on classification performance and enable robust model comparisons. In all cases, human annotation was treated as the ground truth.

- *Precision*, also known as positive predictive value, represents the proportion of correctly predicted positive cases among all cases predicted as positive. In this study, it reflects the probability that tweets classified as critical by an LLM were also labelled as critical by human annotators. A precision score of 100% would indicate that all tweets classified as critical by the model were indeed judged critical by humans, with no false positives.
- *Recall*, also known as sensitivity, denotes the proportion of correctly predicted positive cases among all actual positive cases. Here, it reflects the probability that tweets labelled as critical by human annotators were also classified as critical by the model. A recall score of 100% would indicate that the model successfully captured all tweets deemed critical by humans.
- *F1-Score* combines both precision and recall into a single harmonic mean, offering a balanced measure of classification performance. It is calculated as:

$$F1 - Score = \frac{2 \times Precision \times Recall}{Precision + Recall}$$

The F1-Score ranges from 0 to 1, with higher values indicating better overall performance. In this study, F1-Scores are reported in percentage format for consistency with other metrics. An F1-Score of 100% would signify that the model achieves perfect precision and recall, flawlessly identifying all critical tweets with no misclassifications.

## 4. Results

*4.1 Distribution of critical tweets across retracted and non-retracted articles*

Approximately 6.2% of tweets referencing retracted articles were manually classified as critical of the corresponding research. In contrast, only 1.1% of tweets referencing non-retracted articles were annotated as critical, revealing a substantial disparity in critical discourse between the two groups (Figure 4a).

At the article level, 8.3% of retracted articles were associated with at least one critical tweet prior to retraction, compared to only 1.5% of non-retracted articles (Figure 4b). These results suggest that nearly one in twelve retracted articles in the sample could have been flagged for scrutiny based on pre-retraction Twitter activity. Importantly, the 1.5% of non-retracted articles that attracted critical commentary may also warrant closer scrutiny, as they could potentially harbour unresolved issues.

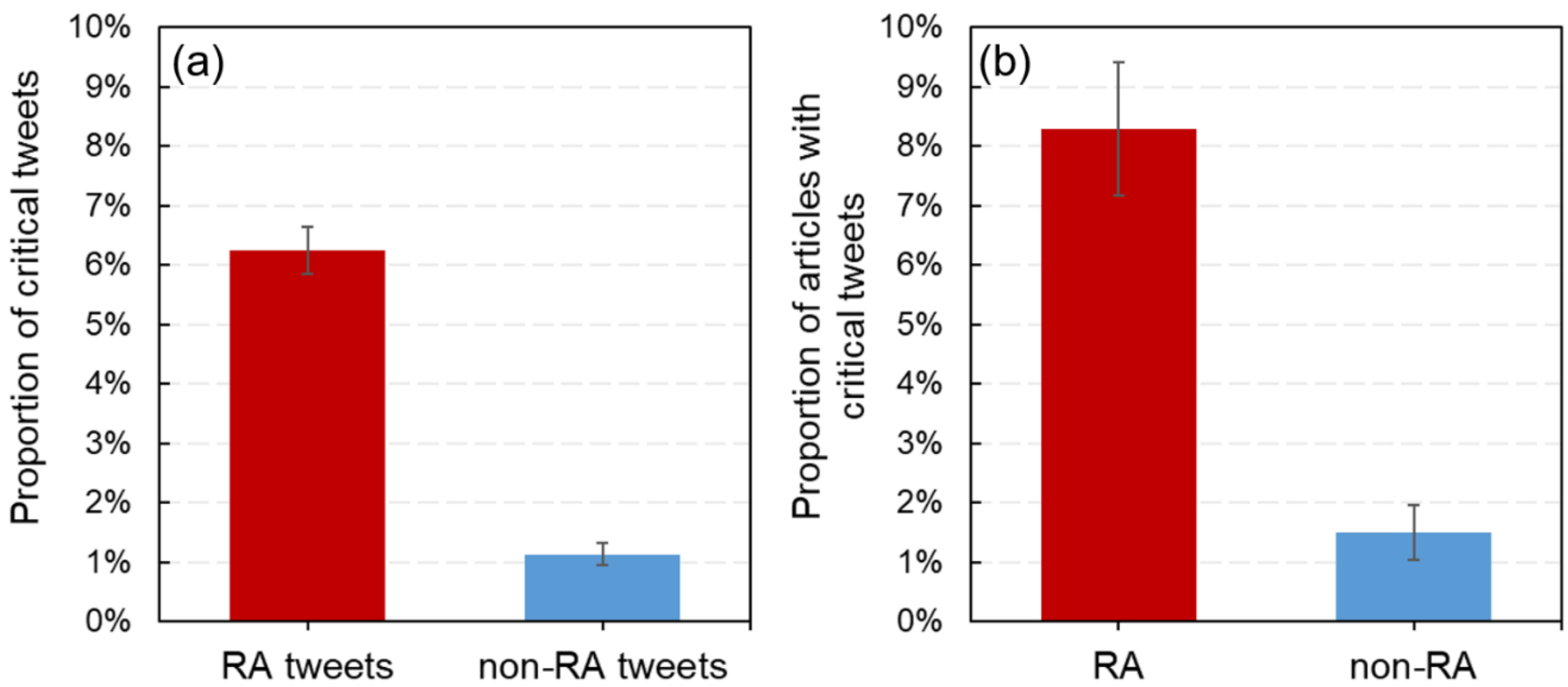

**Figure 4.** (a) Proportion of critical tweets, and (b) proportion of retracted (RA) and non-retracted (non-RA) articles associated with at least one critical tweet.

To explore this pattern further, we plotted a survival function showing the cumulative distribution of critical tweets across articles (Figure 5). The results indicate that 3.6% of retracted articles received two or more critical tweets, and 1.0% received ten or more. In contrast, only 0.6% of non-retracted articles received at least two critical tweets, and just 0.1% received ten or more. These results suggest not only that retracted articles are more likely to receive critical attention overall, but also that they tend to accumulate a greater volume of critical commentary at the individual article level.

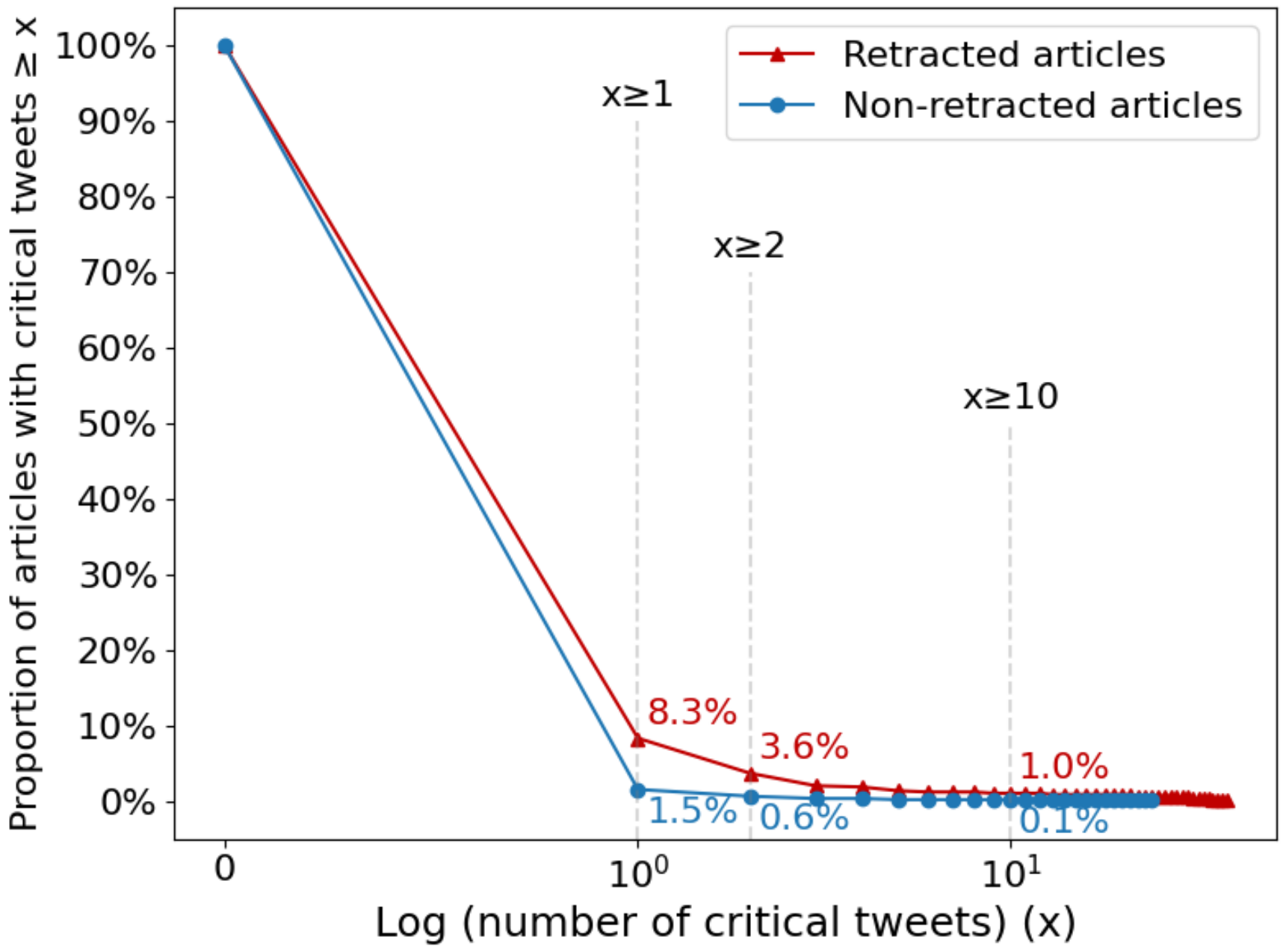

**Figure 5**. Distribution of critical tweets across retracted and non-retracted articles.

*4.2 Comparison of classification results between human annotation and LLMs*

Table 4 presents the performance of each LLM in classifying tweets, using human annotation as the benchmark. Across all LLMs, few-shot prompts consistently yielded superior overall performance compared to zero-shot prompts, as reflected by higher F1-Scores.

**Table 4.** Performance of different LLMs using human annotation as the benchmark.

| Method | Prompt setting | Precision (%) | Recall (%) | F1-Score (%) |
|---|---|---|---|---|
| GPT-4o mini | Zero-shot | 31.33 | 77.54 | 44.63 |
| | Few-shot | 40.24 | 73.91 | 52.11 |
| Gemini 2.0 Flash-Lite | Zero-shot | 49.85 | 61.59 | 55.10 |
| | Few-shot | 48.59 | 68.84 | 56.97 |
| Claude 3.5 Haiku | Zero-shot | 22.10 | 77.17 | 34.36 |
| | Few-shot | 30.12 | 73.55 | 42.74 |
| LLMs consensus | Zero-shot | 34.92 | 74.28 | 47.51 |
| | Few-shot | 43.78 | 73.91 | 54.99 |

However, regardless of the prompt setting, all LLMs exhibited relatively low precision – ranging from 22.10% to 49.85% – indicating a substantial rate of false positives. In other words, many tweets classified as critical by the models were not judged as such by human annotators. In contrast, recall values were generally higher (61.59% to 77.54%), suggesting that the models were moderately effective in capturing tweets considered critical by human annotators.

Nevertheless, the recall scores remained far from prefect, reflecting the limitations of LLMs in reliably detecting critical commentary directed at scholarly articles.

Importantly, the consensus output across the three LLMs did not outperform the best-performing individual model, Gemini 2.0 Flash-Lite, which achieved the highest F1-Score across all configurations. This suggests that simple ensemble voting does not necessarily enhance classification performance in this context.

To better understand the nature of classification errors, we conducted a qualitative analysis of selected false positives, i.e., tweets marked as critical by the LLMs but judged non-critical by human annotators. As shown in Table 5, these misclassifications often stemmed from the model's difficulty in distinguishing tweets that directly critique an article from those that merely reference it. In many instances, tweets used the article as a springboard to discuss broader topics, such as media coverage, ethical debates, or other research, rather than criticising the article itself.

**Table 5**. Examples of tweets misclassified by LLMs as critical of articles, with explanations from human annotators.

| Example tweet | Explanation by human annotators |
|---|---|
| *"Has Jurassic Park become a reality?"* | This expresses astonishment at the attempt described in the article, not criticism. |
| *"Our findings uncovered substantial evidence warranting a closer examination of the research conducted by this team. This evidence raises serious doubts about the reliability of most of their published studies."* | The tweet promotes the authors' own critique of other studies, not the article itself. |
| *"You can find the original research paper via this link. Upon review, it's clear that the actual findings significantly underdeliver compared to the overstated claims in the news."* | The critique targets media misrepresentation, not the article directly. |
| *"I'm no specialist by any means, but it seems like the kind of information you'd want to have in hand prior to experimenting with CRISPR-engineered babies."* | The tweet prompts ethical reflection, not direct criticism of the article's content. |

*Note:* All tweets have been paraphrased to protect user privacy.

## 5. Discussion

### *5.1 The potential of Twitter discussions to support research integrity*

Human annotation revealed that 8.3% of retracted articles were associated with at least one critical tweet prior to retraction, compared to only 1.5% of non-retracted articles. This substantial difference indicates that critical tweets are more commonly associated with articles ultimately retracted. These tweets often highlight methodological flaws, express concerns about misconduct, or employ sarcasm and irony to question an article's credibility. Consequently, critical discourse on social media may serve as an early – albeit partial – indicator of potential problematic research.

Our findings, based on a large, systematically constructed dataset, corroborate insights from earlier small-scale case studies (Haunschild & Bornmann, 2021), reinforcing the value of social media as a supplementary channel for post-publication scrutiny. Twitter thus has potential as a low-cost, real-time early warning system for stakeholders concerned with research integrity. Journal editors could integrate Twitter monitoring into post-publication workflows to identify articles warranting further review. Funding agencies and research integrity offices might use such signals to prioritise investigative resources. For researchers, early critical feedback via social media could reveal methodological weaknesses, limit the dissemination of flawed findings, and strengthen the self-correcting mechanisms of science.

Previous studies have demonstrated that negative sentiment and critical keywords are more prevalent in tweets referencing retracted articles (Amiri et al., 2024; Dambanemuya et al., 2024; Peng et al., 2022). Our manual and LLM-based classifications refine these observations by highlighting that negative sentiment or critical keywords do not necessarily reflect criticism of the article itself. Many tweets direct criticism at related societal issues, media representations, or broader debates, rather than at the article's content or methods. This insight underscores the need to move beyond simplistic sentiment analysis and instead consider the referential focus and rhetorical intent of social media discourse.

Despite the presence of some critical tweets, the majority of tweets about retracted articles are not critical. Several factors may account for this. First, Twitter is not optimally designed for scholarly discussions. Many users are not academics and may lack the expertise to offer critical evaluations (Haustein, 2019; Zhang et al., 2023). Second, scholarly Twitter activity is often geared towards dissemination rather than critique (Didegah et al., 2018; Mohammadi et al., 2018), with many tweets merely sharing article links without substantive commentary (Robinson-Garcia et al., 2017). Third, bot accounts play a notable role in disseminating scientific content on Twitter – both for regular articles (Didegah et al., 2018) and for retracted ones (Dambanemuya et al., 2024) – yet rarely engage in substantive or critical assessments. Together, these factors limit the overall prevalence of critical tweets, even for articles that are eventually retracted.

### *5.2 The potential of LLMs to support research integrity*

Although concerns have arisen regarding the misuse of LLMs in academic writing and data fabrication (Conroy, 2023a; Naddaf, 2023; Silva et al., 2023), these tools offer promising avenues for supporting research evaluation and scientific standards. For example, LLMs have been used to predict future Nobel Prize laureates by identifying major discoveries made by living scientists (Conroy, 2023b), and to generate peer review comments and quality assessments for scholarly articles, often aligning well with human evaluations (Liang et al., 2024). LLM-generated quality scores have also been shown to correlate positively with expert assessments in evaluative contexts (Thelwall et al., 2025).

In more direct applications related to research integrity, LLMs have been used to detect predatory journals (Al-Moghrabi et al., 2024), identify retracted COVID-19 articles (Jan, 2025), and assist with error detection through AI-based tools such as the *Black Spatula Project* and *YesNoError* (Gibney, 2025). Our study extends this emerging literature by exploring whether LLMs can classify tweets as critical of scholarly articles, thereby assessing the feasibility of scalable, automated monitoring of social media as a supplementary mechanism for identifying potential research misconduct.

Despite prior evidence that LLMs can match or exceed human performance in various annotation tasks (Gilardi et al., 2023; Ziems et al., 2024), our findings reveal only partial alignment between LLM-generated labels and human annotations for critical tweets. This suggests that caution is warranted when interpreting the outputs of LLMs in this domain.

Nevertheless, LLMs could be used to triage social media data by filtering out non-critical tweets, directing human attention to potentially concerning posts. Such a hybrid approach could reduce the workload of human reviewers while maintaining high sensitivity. Although misclassifications may still occur, a two-step workflow – initial classification by LLMs followed by human validation – offers a scalable and efficient alternative to human annotation alone. Articles generating a high volume of tweets flagged as critical by LLMs could be prioritised for further investigation, with human reviewers distinguishing tweets that directly critique the article from those addressing unrelated issues.

A key contribution of this study lies in shifting the focus of AI-based detection from article content to the surrounding social media discourse. Instead of solely scrutinising articles, we examined public conversations that may reveal otherwise hidden concerns. This approach expands existing research integrity monitoring strategies and highlights the role of social media not only in dissemination, but also in mobilising collective intelligence to safeguard scientific credibility.

However, several cautions must be highlighted regarding the use of LLMs in this context. First, LLMs are trained on broad and heterogeneous corpora, which may embed social, cultural, or epistemic biases that can affect classification accuracy (Gallegos et al., 2024; Y. Guo, Guo, et al., 2024). Second, deploying advanced models such as GPT-4.5, Gemini 2.5 Pro, or Claude Opus 4 at scale can be computationally intensive and costly (Sathish et al., 2024), raising concerns about the accessibility and scalability of such approaches, particularly for resource-constrained institutions. Third, tweets may contain copyrighted, personal, or sensitive information (Kierkegaard, 2010; Small et al., 2012), and their automated processing by LLMs raises privacy and intellectual property concerns. Ensuring compliance with data protection regulations, such as the General Data Protection Regulation (GDPR)[5], is essential in real-world monitoring scenarios. Finally, while automation can enhance efficiency, overreliance on LLMs may erode critical human judgment, especially in cases requiring contextual understanding or disciplinary expertise. Persistent trust in model outputs could diminish researchers' agency and creativity in making independent critical evaluations (Kumar et al., 2025).

### *5.3 Limitations of the study*

Several limitations should be considered when interpreting the findings of this study. First, although human annotation was used as the benchmark for evaluating LLM performance, human judgment is inherently subjective and susceptible to bias. While the two annotators discussed uncertainties to reach agreement, some misclassifications may remain, potentially influencing the results.

Second, not all critical tweets reflect verifiable issues in the referenced articles, nor do they reliably predict eventual retractions. In some cases, genuine criticism may arise from ideological disagreement rather than methodological flaws. For instance, scientifically rigorous climate research may attract negative responses from climate change denialists. As such, the presence of critical tweets should be interpreted as a preliminary signal rather than a definitive judgment of an article's quality or integrity.

Third, this study did not differentiate between retraction reasons, such as methodological errors, data fabrication, ethical violations, or plagiarism (F. C. Fang et al., 2012). These distinct causes may elicit varying patterns of social media engagement. Future research could investigate whether specific categories of retraction are more readily detectable through critical online discourse.

Fourth, we did not analyse the thematic content of the articles, which likely affects both the frequency and tone of Twitter engagement. Controlling for thematic variation could help clarify the relationship between content and critique. Additionally, our analysis focused exclusively on tweet text and did not incorporate engagement metrics (e.g., likes, retweets, and replies) or contextual data about users (e.g., whether the account belongs to a bot, journalist, or academic). These factors may shape both the interpretation and amplification of tweets and should be considered in future work (Dambanemuya et al., 2024).

Lastly, recent structural changes to Twitter – such as the shift to paid API access and stricter rate limits – have constrained researchers' ability to collect large-scale, up-to-date datasets. These changes may also influence the volume and nature of academic discourse on the platform. In parallel, many researchers are migrating to alternative platforms such as Bluesky or Mastodon (Bittermann et al., 2023; Kupferschmidt, 2024; Mallapaty, 2024; Vidal Valero, 2023), reducing Twitter's representativeness as a venue for science communication. Nonetheless, the methodological framework developed in this study is adaptable and can be applied to other platforms. This presents opportunities for future research to extend the investigation of scholarly discourse and critical engagement across a broader social media ecosystem.

[5] For more information on this regulation, see: https://gdpr-info.eu/ (accessed on 24 June 2025).

## 6. Conclusions

This study examined whether critical tweets can serve as early warning signals for retracted scholarly articles and assessed the potential of large language models (LLMs) to automate the detection of such tweets. Manual analysis revealed that 8.3% of retracted articles were associated with at least one critical tweet prior to retraction, compared to just 1.5% of non-retracted articles. These findings underscore Twitter' potential as a supplementary channel for post-publication review and research integrity monitoring.

LLMs' ability to detect critical content aligned only moderately with human annotation. In particular, the models' low precision indicates a relatively high rate of false positives. This suggest that the selected LLMs, in their current form, are not yet reliable as standalone tools for critical tweet detection. However, they hold promise as components of hybrid workflows that combine automated filtering with human validation to support scalable monitoring.

This study contributes a novel perspective by shifting attention from analysing article content to examining surrounding social media discourse. Despite certain limitations – including the relative rarity of critical tweets and the evolving platform constraints – our methodological framework offers a scalable, transferable approach for identifying potentially problematic research through real-time public commentary. Future work could extend this framework to emerging platforms and incorporate additional contextual signals to improve detection accuracy and enhance practical utility.

## Acknowledgments

Zhichao Fang is financially supported by the National Natural Science Foundation of China (No. 72304274). Hui-Zhen Fu is supported by the National Social Science Foundation of China (No. 22CTQ032). Er-Te Zheng is financially supported by the GTA scholarship from the School of Information, Journalism and Communication of the University of Sheffield. Mike Thelwall is supported by the Fundação Calouste Gulbenkian European Media and Information Fund (No. 316177). The authors thank Altmetric and Retraction Watch for providing the data for research purposes, and the anonymous reviewers for their valuable comments.